\documentclass{Interspeech}

\interspeechcameraready

\title{Cross-lingual Data Selection Using Clip-level Acoustic Similarity for Enhancing Low-resource Automatic Speech Recognition}

\author[affiliation={1}]{Shunsuke}{Mitsumori}
\author[affiliation={1}]{Sara}{Kashiwagi}
\author[affiliation={2}]{Keitaro}{Tanaka}
\author[affiliation={2}]{Shigeo}{Morishima}

\affiliation{}{Waseda University}{Japan}
\affiliation{}{Waseda Research Institute for Science and Engineering}{Japan}
\email{mitsumori@ruri.waseda.jp, sara.kashiwagi@moegi.waseda.jp,\\ keitaro@aoni.waseda.jp, 	shigeo@waseda.jp}
\keywords{automatic speech recognition, cross-lingual, low-resource language, deep learning}

\usepackage{comment}
\usepackage{xcolor}
\usepackage{cite}
\usepackage{adjustbox}

\newcommand{\tkrevdone}[1]{\textcolor{black}{#1}}

\begin{document}

\maketitle

\begin{abstract}
This paper presents a novel donor data selection method to enhance low-resource automatic speech recognition (ASR). 
While ASR performs well in high-resource languages, its accuracy declines in low-resource settings due to limited training data. 
A common solution is to leverage multilingual self-supervised learning (SSL) models with donor languages. 
However, existing methods rely on language-level similarity, overlooking clip-level variations. 
To address this limitation, we propose clip-wise acoustic token distribution similarity (CATDS), a fine-grained selection method that identifies acoustically relevant donor clips for better alignment with the target language. 
Unlike existing clip-level selection methods, our method aligns with the representation of SSL models and offers more challenging yet valuable samples. 
Experimental results show that CATDS outperforms traditional selection methods and can even utilize donor languages previously considered detrimental.
    
    
    
\end{abstract}


\section{Introduction}
\tkrevdone{Automatic speech recognition (ASR) has become a foundational technology for converting spoken language into text.
It has widespread applications, including smartphone assistants~\cite{macoskey21b_interspeech,ding22_interspeech,10.1145/3479532}, meeting transcription~\cite{kanda21_interspeech,1198793,156581}, and broadcast captioning~\cite{alumae-etal-2023-automatic,prazak21_interspeech,10.1145/3613904.3642177}.
ASR systems have achieved remarkable success in well-resourced languages by demonstrating high accuracy, but their effectiveness diminishes in low-resource languages~\cite{khare21_interspeech,9415020,JMLR:v25:23-1318}.
This is primarily due to the lack of training data, which deepens the technological divide.}

\tkrevdone{A major approach to addressing this low-resource issue is leveraging multilingual self-supervised learning (SSL) models, such as XLSR~\cite{conneau21_interspeech,babu22_interspeech}. 
By simultaneously learning from multiple languages, these SSL models can capture shared representations that transfer to low-resource scenarios, leading to notable accuracy gains. 
Moreover, prior studies have shown that identifying linguistically similar languages to a low-resource target language 
 and incorporating them as \textit{donors} during pre-training can further improve ASR accuracy~\cite{adams-etal-2019-massively,atds}.}

\tkrevdone{The linguistic similarity of individual clips to the target language, however, can vary significantly across utterances or sentences, even within a single donor language.
Recent studies have highlighted the importance of prioritizing clip-level resemblance over language-level similarity~\cite{10848811,cheng2024exploringimpactdataquantity}.
To this end, existing research has explored language identification (LID) models to assess resemblance
 by estimating the likelihood that a given clip belongs to the target language.
Nonetheless, given the substantial differences between LID models and SSL-based cross-lingual ASR models (e.g., XLSR) 
 in task, architecture, and training methodology, 
 repurposing LID-based features to quantify the contribution of each donor clip to target ASR training 
 appears to be a suboptimal strategy.}

\tkrevdone{In this paper, we present a novel method for donor clip selection that leverages acoustic feature similarity.
LID models are trained to be discriminative in classification tasks, even distinguishing acoustically similar clips across different languages and categorizing them into distinct classes.
In contrast, SSL models learn phonetic characteristics across languages in a self-organized manner~\cite{choi24b_interspeech} 
 and can map these clips to similar representations in a shared latent space.
As a result, the selected donor clips, based on the acoustic similarity, are expected to serve as challenging yet valuable samples
 that can promote efficient training and improve the performance of eventual SSL-based cross-lingual ASR models.}

\tkrevdone{To facilitate data selection, we propose clip-wise acoustic token distribution similarity (CATDS), which advances the language-level acoustic token distribution similarity (ATDS)~\cite{atds} to a finer granularity.
Our approach first converts donor speech clip candidates into sequences of pseudo acoustic tokens using the wav2seq~\cite{wav2seq} pipeline,
 encoding audio frames into token representations.
We then compute the frequency distribution of these tokens and represent them as frequency vectors. 
Next, we measure the cosine similarity between each clip's frequency vector and the language-level frequency vector of the target language.
Because similarity scores inadvertently correlate with the total number of tokens in a clip, 
 we apply a scaling adjustment to derive the final CATDS value. 
By selecting the top-ranked clips based on CATDS, our method constructs a refined donor dataset 
for SSL model training.}


The main contribution of this work is the development of CATDS, a new data selection method capable of identifying beneficial clips from donor datasets for improved low-resource ASR.
Unlike conventional strategies that rely on LID models, our approach employs the same SSL model used for ASR 
 during data selection.
This ensures consistency between the selection and training phases, maintaining alignment in the representation space and fostering more efficient training.
Experimental results demonstrate that our method outperforms baseline approaches, including the existing LID-based method.
Notably, for a donor language found to be beneficial overall, 
 our method achieved higher accuracy by discarding 40\% of low-CATDS clips than by using the entire dataset.
Furthermore, even for donor languages that negatively impacted recognition performance when used in full, 
 our method improved results by selecting only the top 20\% of clips ranked by CATDS.
This outcome suggests that data previously considered detrimental can now be effectively leveraged, paving the way for addressing general low-resource challenges in ASR. Our code is available at \url{https://github.com/Shunsuke32/CATDS.git}.

\begin{figure*}[t]
    \centering
    \includegraphics[width=\textwidth]{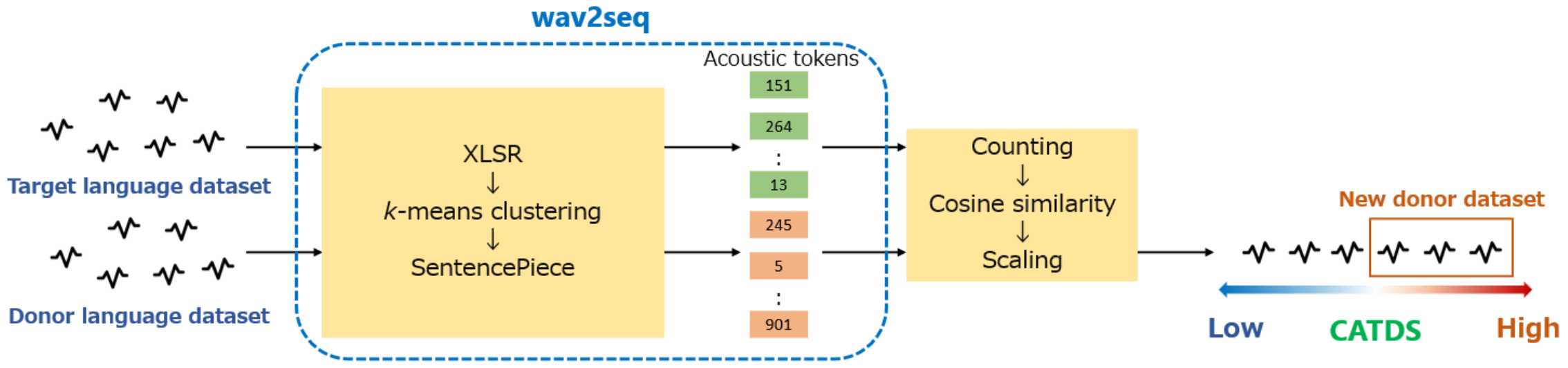}
    \caption{Our method introduces clip-wise acoustic token distribution similarity (CATDS), which selects donor language data based on acoustic similarity to enhance low-resource ASR.}
    \label{fig:method} 
\end{figure*}

\section{Related Work}
This section briefly reviews existing works on language-level acoustic similarity and clip-level data selection methods.

\subsection{Language-level acoustic similarity}
Nay San et al.~\cite{atds} developed a similarity metric called ATDS to identify donor languages that enhance ASR performance when included in training for a given target language. Experimental results demonstrated that the ATDS values exhibit a high correlation with improvements in the target language’s accuracy, surpassing the correlation achieved by an existing approach based on the LID model. This outcome underscores the effectiveness of the ATDS-based language selection method. Furthermore, the proposed approach extends the ATDS framework from the language level to the clip level, enabling the construction of more effective datasets for target language ASR training.

\subsection{Clip-level data selection}
Chen et al.~\cite{10848811} proposed a method that uses only those clips deemed similar to the target language based on an LID model~\cite{speechbrain} selected from the entire dataset of a closely related language. In practice, when the LID model identifies the language of a non-target clip, this approach uses the rank of the target language among the predicted possibilities (a top-k scheme) as a measure of similarity. Their experimental results showed that ASR training on the dataset selected by this method achieved higher accuracy than randomly selecting data.

In contrast, our proposed approach aims to measure acoustic similarity more closely tied to the SSL model used in training, rather than relying on an LID model for feature extraction. By utilizing the same SSL model that will be trained for ASR, we seek to ensure the selected data align more effectively with the model’s learned representations.

\section{Proposed Method}
In this study, we extend the ATDS method~\cite{atds} to measure acoustic similarity at the clip level (Figure~\ref{fig:method}).
We first explain the method for calculating acoustic token frequency distributions and then introduce a metric to evaluate the similarity of individual donor clips to the target language.

\subsection{Tokenization of speech data using wav2seq}
First, we segment the speech data into 20~ms frames and input them into the pre-trained XLSR model to extract embedding sequences.
Specifically, we obtain 1024-dimensional intermediate features from the 12th layer of XLSR's 24-layer Transformer encoder.
This choice is motivated by prior research demonstrating that intermediate layers in speech processing models effectively capture acoustic characteristics \cite{2103.14583,BARTELDS2022101137,10095033,bartelds-wieling-2022-quantifying,guillaume23_sigul}.

Next, we apply \textit{k}-means clustering to these intermediate feature sequences, grouping feature vectors into clusters.
We set the number of clusters (i.e., $k$) to 500 and train the clustering model using a 5-hour subset of the target language data.
Each cluster index is then mapped to a character (e.g., 10 $\rightarrow$ `k', 19 $\rightarrow$ `t'), enabling compatibility with SentencePiece~\cite{sentencepiece}. This mapping allows the cluster indices to function as phoneme-like categories, making them suitable for tokenization.

Afterward, consecutive duplicate phoneme categories are removed to reduce redundancy.
The resulting sequences are then tokenized using the SentencePiece model (vocabulary size $V=10000$),
 which has been trained on the same 5-hour subset of the target language data.
This process generates a sequence of acoustic tokens that effectively capture frequently occurring subword acoustic units.

Finally, we construct token frequency distributions for each speech clip by counting the occurrences of individual tokens.
Formally, each clip's token sequence is mapped to a $V$-dimensional frequency vector, where $V$ corresponds to the SentencePiece vocabulary size (i.e., 10000).
These frequency distributions serve as the basis for quantifying clip-level acoustic similarities, 
 enabling a more fine-grained assessment of phonetic resemblance across clips.

\subsection{Token frequency distribution similarity calculation}
To assess the similarity of each donor clip to the target language, we compare its token frequency distribution with that of the 5-hour target language subset. 
Let $\mathbf{x}$ denote the frequency vector of the target language subset and $\mathbf{y}$ the frequency vector of a given donor clip. We compute their cosine similarity as
\begin{align}
    S(\mathbf{x}, \mathbf{y}) = \frac{\mathbf{x} \cdot \mathbf{y}}{\lVert \mathbf{x} \rVert \lVert \mathbf{y} \rVert}.
    \label{eq:cosine}
\end{align}

\begin{figure}[t]
    \centering
    \includegraphics[width=\columnwidth]{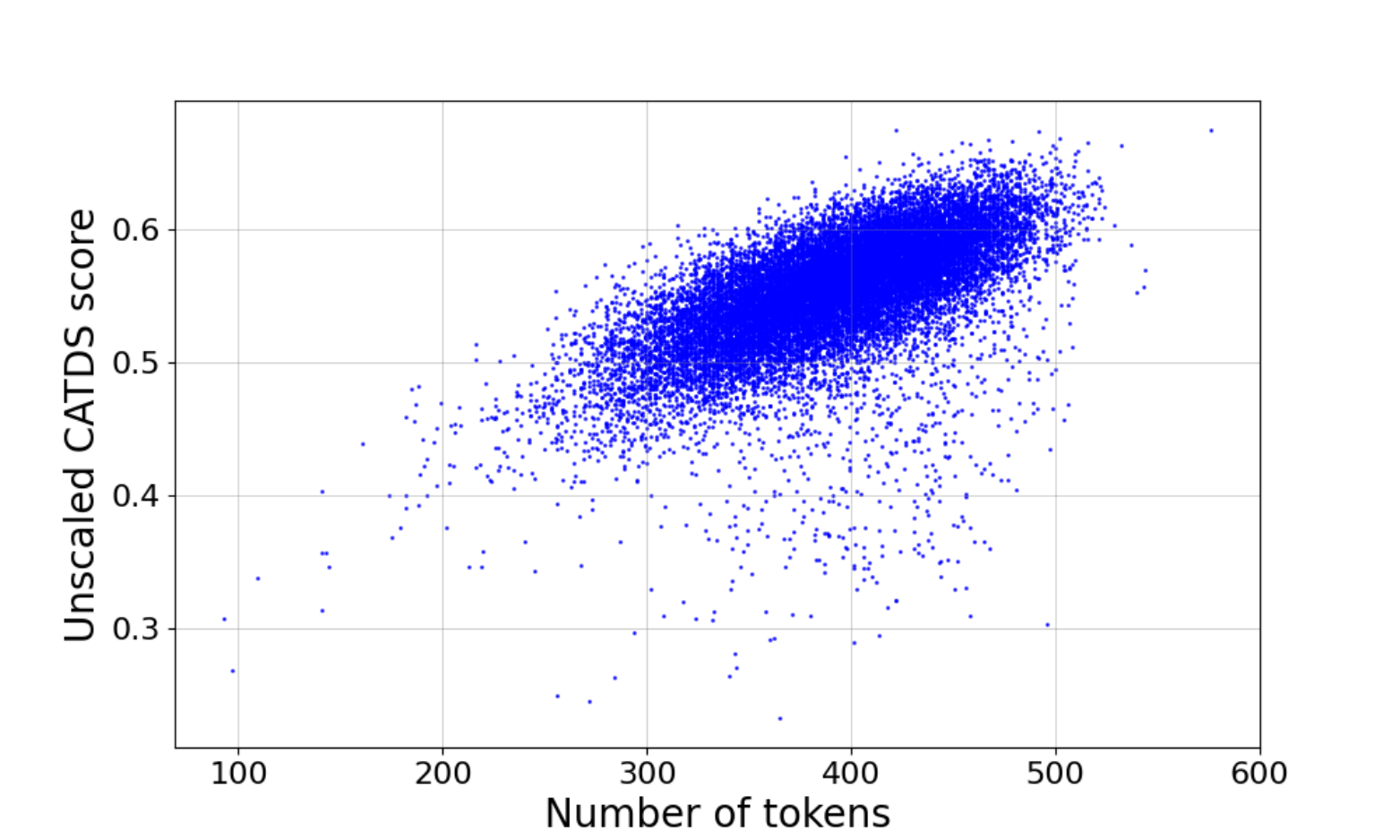}
    \centering
    \includegraphics[width=\columnwidth]{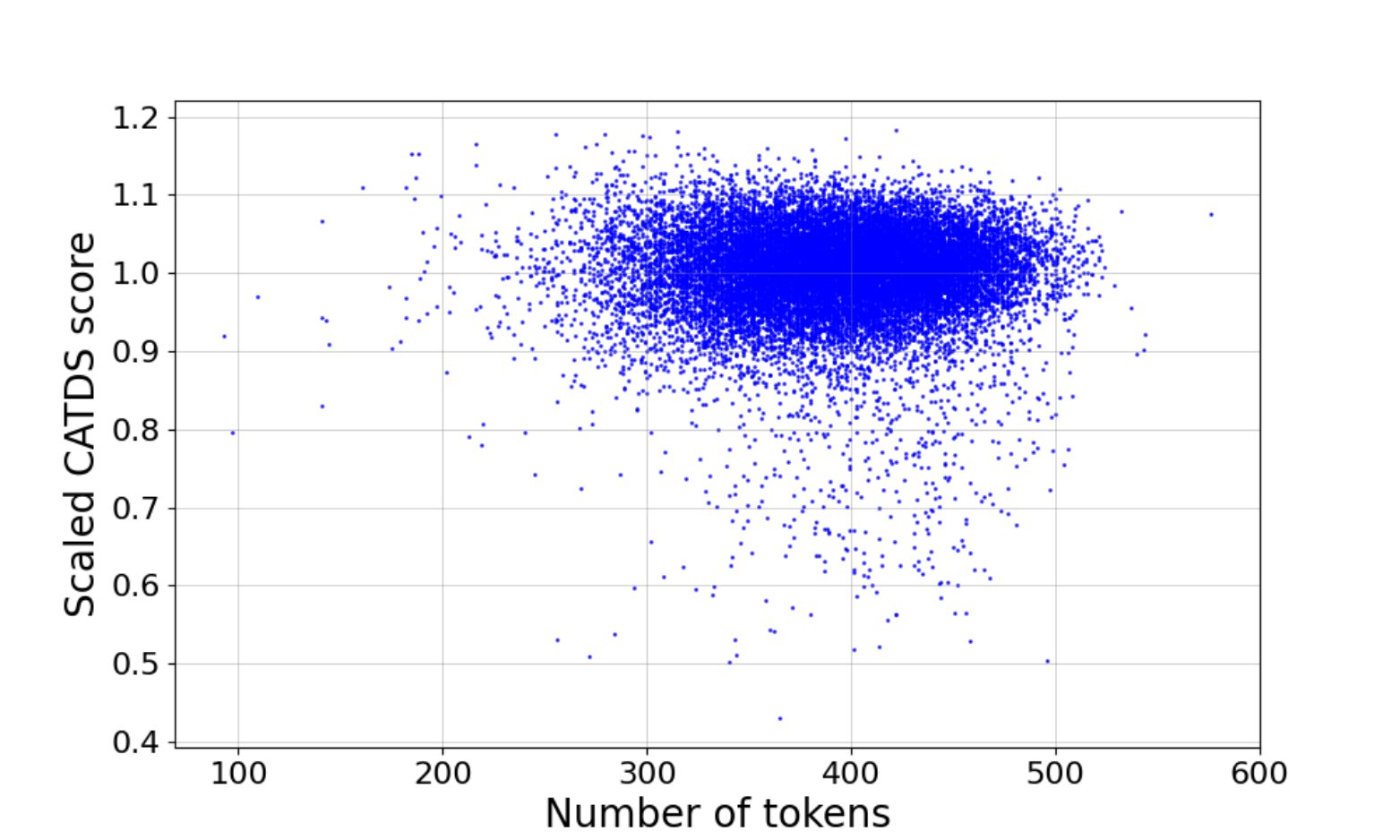}
   \caption{ Relationship between the number of tokens and the CATDS score before scaling (top) and after scaling (bottom).}
   \label{fig:before_after}
\end{figure}

However, raw cosine similarity tends to increase with clip length, as longer clips contain a larger number of tokens (the upper part of Figure~\ref{fig:before_after}). Since our goal is to measure acoustic similarity rather than clip duration, we apply a quadratic regression model for scaling adjustment to account for this dependency:
\begin{align}
    q = a p^2 + b p + c,
    \label{eq:quad}
\end{align}
where $p$ is the clip’s token count, $q$ is the similarity value predicted by the quadratic regression model fitted to the raw cosine similarity as a function of $p$, and $a$, $b$, and $c$ are the learnable regression coefficients.
The final CATDS value is then obtained by scaling the raw similarity $S$ by the fitted value $q$,
\begin{align}
    \mathrm{CATDS} = \frac{S}{q}.
    \label{eq:catds}
\end{align}

The bottom part of Figure~\ref{fig:before_after} illustrates the effect of this scaling adjustment. 
While the unscaled similarity increases with the number of tokens, the scaled CATDS metric shows no clear dependence on token count. This demonstrates that our scaling approach effectively eliminates length bias and ensures that CATDS accurately reflects inherent acoustic similarity.

\section{Evaluation}
This section describes experiments conducted to evaluate the performance of the proposed CATDS-based data selection method for low-resource ASR.

\subsection{Dataset}
In our experiments, we used the IndicSUPERB dataset~\cite{IndicSUPERB}, which contains speech data from 12 Indian languages.
Since Punjabi is a representative low-resource language~\cite{atds}, we selected it as the target language.
To investigate the impact of different donor languages on Punjabi speech recognition, we used Hindi, Malayalam, and Bengali during training.
We conducted experiments with three configurations: (Punjabi, Hindi), (Punjabi, Malayalam), and (Punjabi, Bengali).

The training process follows a two-stage framework that consists of continued pre-training~\cite{NOWAKOWSKI2023103148,10301554} and fine-tuning, following the XLSR method~\cite{conneau21_interspeech,babu22_interspeech}.
To construct the training sets for continued pre-training, multiple donor samples were concatenated into clips with a maximum duration of 21.6 seconds, resulting in 20,000 clips.
Additionally, to enhance training stability, we incorporated 10 hours of the target Punjabi data into the training sets.
For fine-tuning, we used only Punjabi data, allocating 1 hour for training, 1 hour for validation, and 2 hours for testing.




\subsection{Data selection}
To evaluate the effectiveness of our data selection approaches, we systematically reduce the number of donor language clips from 
\(N = 20{,}000\)
to lower counts in increments of 
\(\Delta N = 4{,}000\).
For each subset size \(S_K\), we define it as follows:
\begin{align}
    S_K &= N - k \cdot \Delta N, 
    \quad 
    k \in \{0, 1, 2, 3, 4, 5\}.
\end{align}
We construct training datasets under three selection strategies.

\noindent \textbf{Random selection.}  
Let \(\mathcal{D}_{\text{donor}}\) be the full donor dataset. We define the randomly selected subset \(\mathcal{D}_{\text{random}, k}\) by
\begin{align}
    \mathcal{D}_{\text{random}, k} 
    \subset 
    \mathcal{D}_{\text{donor}}, 
    \quad 
    \lvert\mathcal{D}_{\text{random}, k}\rvert 
    = S_K.
\end{align}
To mitigate stochastic variation, we generate three independent subsets for each \(S_K\), denoted as 
$\mathcal{D}_{\text{random}, k}^{(i)}, i \in \{1,2,3\}$.

\noindent \textbf{LID-based selection.}  
As a second selection method, we employ a SpeechBrain-based~\cite{speechbrain} LID model~\cite{10848811}.
For each clip \(x\in\mathcal{D}_{\text{donor}}\), let $R(x) \in \{1,2,3,\dots\}$
be the rank of Punjabi among the model’s predicted languages (i.e., \(R(x)=1\) if Punjabi is most likely, \(R(x)=2\) if second most likely, etc.), and $P(x) \in [0,1]$ denote the predicted probability that clip \(x\) is Punjabi. We sort \(\mathcal{D}_{\text{donor}}\) in ascending order of \(R(x)\), and for ties (clips sharing the same \(R(x)\)), we sort in descending order of \(P(x)\). Let 
\(\pi_{\text{LID}}(1),\pi_{\text{LID}}(2),\dots\)
denote the resulting sequence of clips. The LID based subset of size \(S_K\) is then described as:
\begin{align}
    \mathcal{D}_{\text{LID},k}
    =
    \bigl\{
      \pi_{\text{LID}}(i) \mid 1 \leq i \leq S_K
    \bigr\}.
\end{align}

\noindent \textbf{CATDS-based selection.}  
Let \( f_{\text{CATDS}}: \mathcal{A} \to \mathbb{R} \)  
be the CATDS scoring function, where \(\mathcal{A}\) represents an audio clip. The function \( f_{\text{CATDS}}(x) \) assigns a relevance score to each clip \( x \in \mathcal{D}_{\text{donor}} \), indicating its importance for adaptation to the target language.
We sort \(\mathcal{D}_{\text{donor}}\) in descending order of \( f_{\text{CATDS}}(x) \), such that clips with the highest CATDS scores are ranked first. Let  
\(\pi_{\text{CATDS}}(1), \pi_{\text{CATDS}}(2), \dots\)  
denote the resulting ordered sequence of clips. The CATDS based subset of size \( S_K \) is then defined as  
\begin{align}
    \mathcal{D}_{\text{CATDS}, k} =
    \bigl\{
        \pi_{\text{CATDS}}(i) \mid 1 \leq i \leq S_K
    \bigr\}.
\end{align}


By construction, when \(k = 0\) and \(k = 5\), all four selection methods yield identical subsets: at \(k=0\), the dataset contains all 20,000 clips; at \(k=5\), it is empty. Hence, these two extremes produce two \emph{shared} datasets. For each of the intermediate values \(k \in \{1,2,3,4\}\), we create:
\begin{itemize}
    \item Three independent \textbf{random} subsets
    \(\mathcal{D}_{\text{rand}, k}^{(i)}\),
    \(\,i \in \{1,2,3\}\).
    \item One \textbf{LID} based subset
    \(\mathcal{D}_{\text{LID},k}\).
    \item One \textbf{CATDS} based subset
    \(\mathcal{D}_{\text{CATDS},k}\).
    \item One \textbf{Unscaled CATDS} based subset
    \(\mathcal{D}_{\text{uCATDS},k}\).
\end{itemize}
Therefore, for each donor language, we train on a total of 
$ 2\ (\text{shared}) 
  + 
  4 \times \bigl(3\ (\text{random}) + 1\ (\text{LID}) + 1\ (\text{CATDS}) + 1\ (\text{Unscaled CATDS})\bigr) 
  = 26
$
distinct dataset configurations. After continued pre-training on each subset, the model is finetuned \emph{only} on Punjabi to evaluate the impact of different donor data selection strategies.

During continued pre-training, we utilize one hour of Punjabi data as the validation set.
For finetuning, we employ only Punjabi speech data. 
We allocate one hour of Punjabi speech for training, one hour for validation, and two hours for testing.




{\tabcolsep=1.8mm
\begin{table*}[!t]
    \centering
    \caption{\textit{Performance comparison of CATDS (unscaled vs. scaled) across languages and clip numbers.}}
    \vspace{-1.5mm}
    \label{tab:data_selection}
    \begin{tabular}{lcccccccccccc}
        \toprule
        \textbf{Language} & \multicolumn{4}{c}{\textbf{Hindi}} & \multicolumn{4}{c}{\textbf{Malayalam}} & \multicolumn{4}{c}{\textbf{Bengali}} \\
        \cmidrule(lr){2-5} \cmidrule(lr){6-9} \cmidrule(lr){10-13}
        \textbf{Number of clips} & 4000 & 8000 & 12000 & 16000 & 4000 & 8000 & 12000 & 16000 & 4000 & 8000 & 12000 & 16000 \\
        \midrule
        \textbf{Data selection method} \\
        Random & \underline{26.95} & 26.90 & \underline{25.89} & 25.56 & 29.02 & \underline{28.97} & \underline{29.42} & 29.69 & 28.44 & 29.00 & 28.41 & \underline{28.97} \\
        CATDS (unscaled) & 27.25 & \textbf{26.00} & 26.12 & \textbf{25.14} & \underline{28.66} & 29.41 & 29.87 & \underline{29.54} & \textbf{27.86} & \textbf{28.56} & \underline{28.38} & 29.54 \\
        CATDS (scaled) & \textbf{26.74} & \underline{26.60} & \textbf{25.20} & \underline{25.26} & \textbf{28.03} & \textbf{28.96} & \textbf{29.21} & \textbf{29.52} & \underline{28.02} & \underline{28.66} & \textbf{28.14} & \textbf{28.27} \\
        \bottomrule
    \end{tabular}
\end{table*}
}

\begin{figure}[!t]
   \centering
   \textbf{(i) Hindi} \\
   \includegraphics[width=.99\columnwidth]{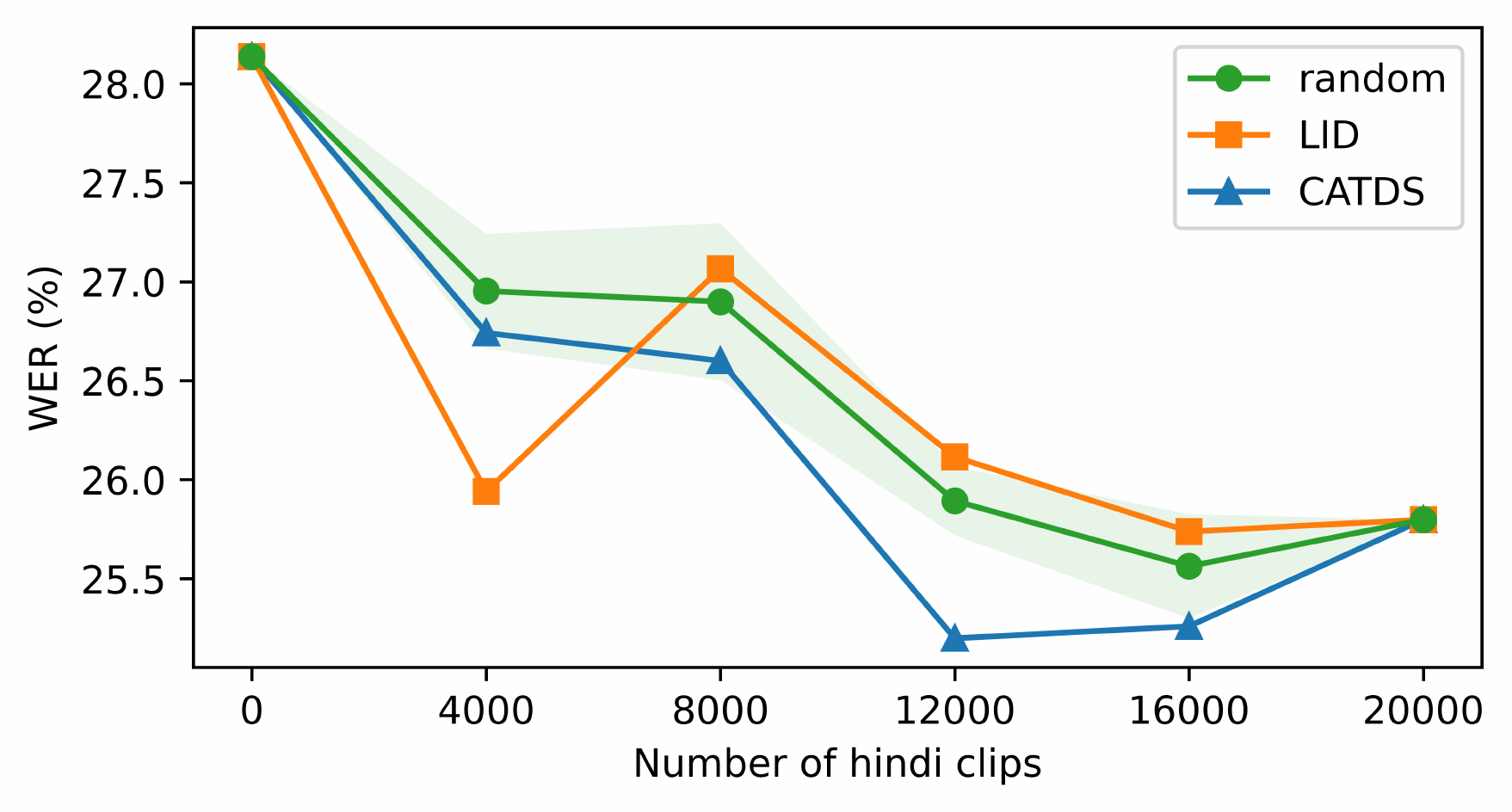}
   \vspace{2mm} 
   
   \textbf{(ii) Malayalam} \\
   \includegraphics[width=.99\columnwidth]{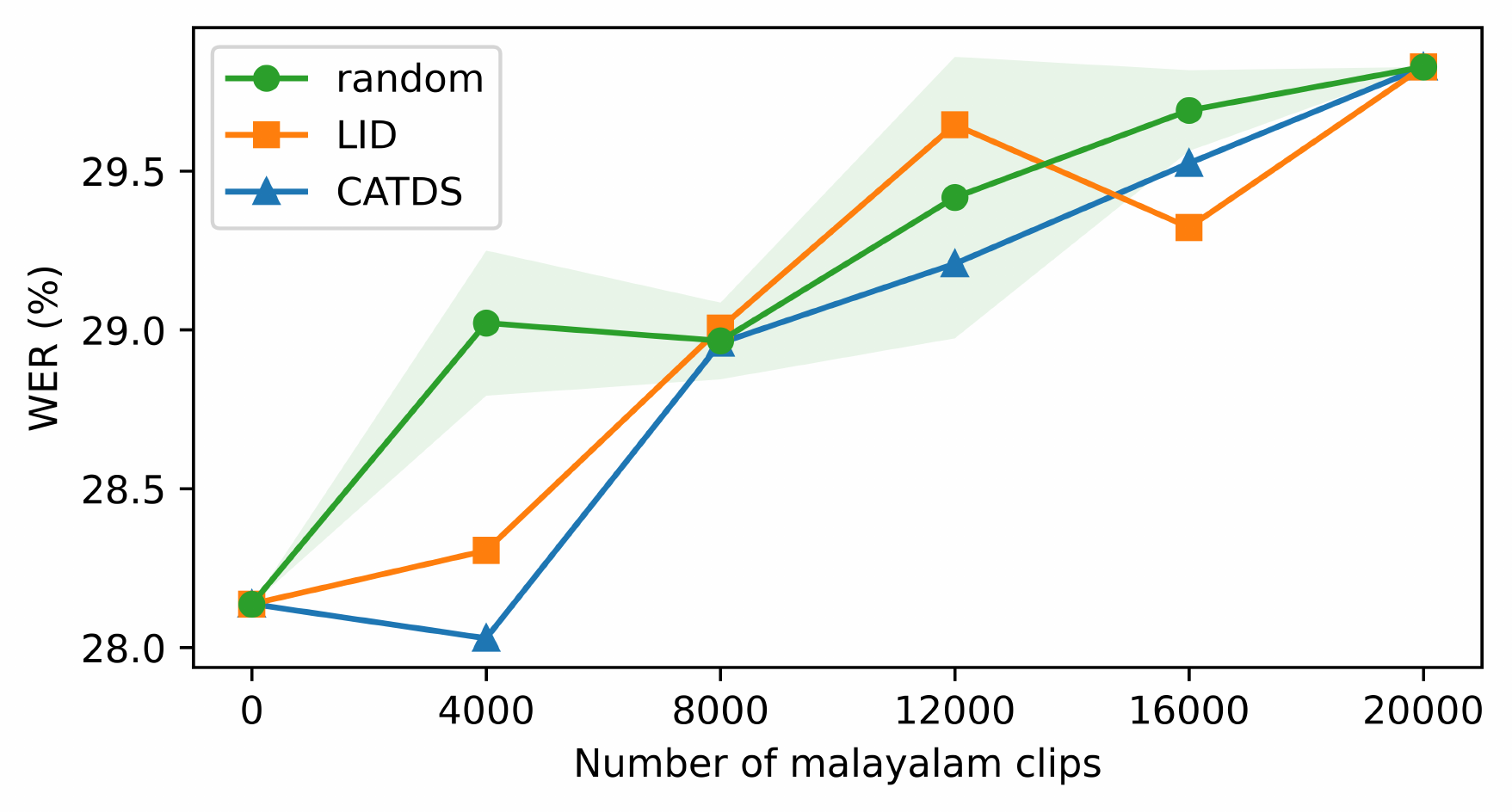}
   \vspace{2mm} 
   
   \textbf{(iii) Bengali} \\
   \includegraphics[width=.99\columnwidth]{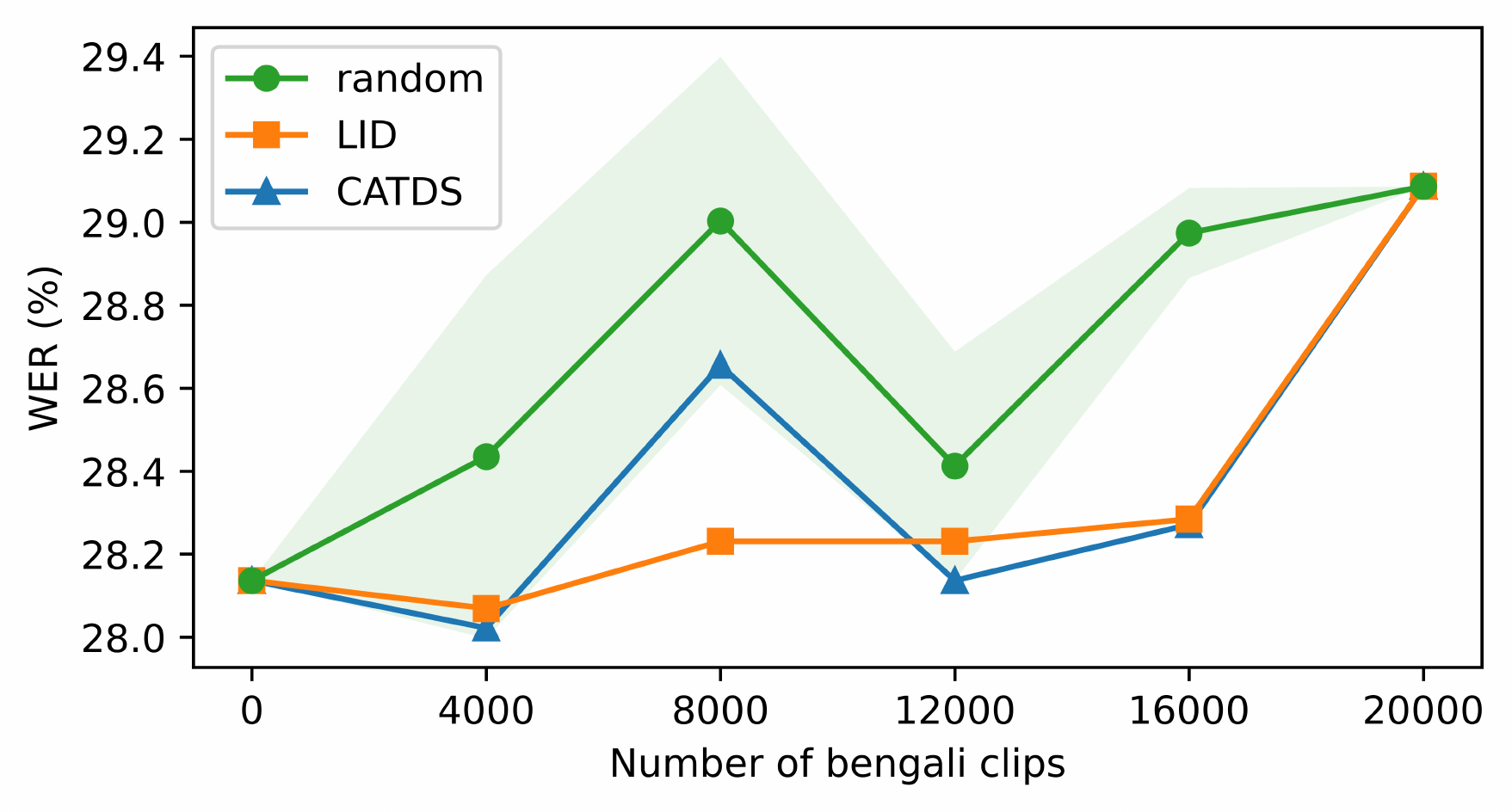}
   \vspace{-5pt}
   \caption{Changes in ASR accuracy with reduced (i)~Hindi (ii)~Malayalam (iii)~Bengali data, from 20000 to 0 by 4000.The shaded area represents the range $m\pm\sigma$, where $m$ is the mean and $\sigma$ is the standard deviation of WER over three experiments.}
   \label{fig:all_langs}
   \vspace{-5mm}
\end{figure}

\subsection{Experimental results}

We conducted experiments using the random, LID-based, and proposed CATDS-based methods, and the results are presented in Figure \ref{fig:all_langs}.
Our proposed method consistently outperformed the random selection across all donor languages.
Specifically, when using Hindi as the donor language, the lowest word error rate (WER) was achieved with 12,000 clips selected by CATDS, demonstrating that the propoesed method successfully excluded clips that negatively impacted the learning of Punjabi.
Unlike Hindi, using the entire dataset from Malayalam and Bengali resulted in lower Punjabi ASR accuracy compared to when it was not used.
However, the lowest WER was recorded with 4,000 clips selected by CATDS, demonstrating that the propoesed method was able to extract only the beneficial clips, mitigating the negative impact and improving recognition accuracy.
While the proposed method underperformed LID-based selection method in certain settings, LID-based selection exhibited lower accuracy than random selection in multiple cases.
This suggests that the proposed method provides a more stable and effective data selection compared to LID-based selection.
To statistically validate the observed advantage of our method, we conducted three paired Wilcoxon signed-rank tests (two-tailed, $n = 12$) comparing each selection method with the random method.  
At the 0.01 significance level, only the CATDS-based method reached significance ($p = 0.00049$), whereas neither the LID-based method ($p = 0.092$) nor the unscaled CATDS-based method ($p = 0.733$) did, underscoring the superiority of our approach.

Figure \ref{fig:before_after} illustrates how scaling affects the relationship between token count and CATDS score.
Before scaling, CATDS scores tended to increase with token count, whereas after scaling, this dependency was effectively eliminated.
The impact of scaling on ASR performance is summarized in Table \ref{tab:data_selection}. 
While unscaled CATDS achieved high accuracy in some settings, it performed worse than random selection in others.
In contrast, scaled CATDS consistently outperformed random selection in all conditions and achieved the best results in most cases.
This confirms that scaling effectively removes token count dependency, allowing CATDS to measure intrinsic acoustic similarity regardless of the number of tokens in a clip.

\section{Conclusion}

In this study, we proposed CATDS, a novel donor data selection method based on clip-level acoustic similarity, to enhance ASR performance for low-resource languages. 
Unlike conventional LID-based methods, the proposed method leverages the same SSL model used for ASR to extract acoustic features, ensuring more consistent data selection.
Experimental results confirmed that CATDS consistently outperforms random selection and achieves comparable or superior performance to the existing LID-based method. Furthermore, for all donor languages, CATDS-selected datasets achieved the lowest WER, demonstrating the effectiveness and stability of CATDS as a data selection method. Even for donor languages (Malayalam, Bengali) that generally degrade ASR performance when the entire dataset is used, training with only the 4,000 CATDS-selected clips successfully improved model accuracy.
This finding suggests that CATDS enables the effective utilization of previously detrimental donor language data, offering new possibilities for improving ASR in low-resource scenarios.

\clearpage






\bibliographystyle{IEEEtran}
\bibliography{mybib}

\end{document}